\documentclass[12pt]{amsart}

\usepackage[foot]{amsaddr}
\usepackage{graphicx}
\usepackage{mathptmx} 
\usepackage{amssymb}
\usepackage{amsmath}
\usepackage{amsthm}
\usepackage{mathtools}
\usepackage[dvipsnames]{xcolor}
\usepackage[sort,round]{natbib}
\usepackage{marvosym}
\usepackage{a4wide}
\usepackage[normalem]{ulem}
\newcommand{\stkout}[1]{\ifmmode\text{\sout{\ensuremath{#1}}}\else\sout{#1}\fi}
\usepackage{enumerate}
\usepackage{url}
\usepackage{csquotes}

\usepackage[mathscr]{euscript}

\theoremstyle{plain}
\newtheorem{theorem}{Theorem}[section]
\newtheorem{lemma}[theorem]{Lemma}

\newtheorem{fact}[theorem]{Fact}

\theoremstyle{definition}
\newtheorem{example}[theorem]{Example}

\newcommand{\supp}{{\rm supp}}

\newcommand{\pf}{\noindent{\em Proof: }}
\newcommand{\epf}{\hfill\hbox{\rule{3pt}{6pt}}\\}

\title{Spaces of ranked tree-child networks}
\author{Vincent Moulton}
\address[V. Moulton]{School of Computing Sciences, University of East Anglia, UK}
\email{v.moulton@uea.ac.uk}
\author{Andreas Spillner}
\address[A. Spillner]{Merseburg University of Applied Sciences, Germany}
\email{andreas.spillner@hs-merseburg.de}


\date{\today}

\begin{document}

\begin{abstract}
Ranked tree-child networks are a recently introduced class
of rooted phylogenetic networks 
in which the evolutionary events represented by the network
are ordered so as to respect the flow of time. This class includes
the well-studied ranked phylogenetic trees (also
known as ranked genealogies). An important problem
in phylogenetic analysis is to define distances
between phylogenetic trees and networks in order to systematically compare them.
Various distances have been defined on ranked binary phylogenetic trees,
but very little is known about comparing ranked tree-child networks.
In this paper, we introduce an approach to compare binary
ranked tree-child networks on the same leaf set that is 
based on a new encoding of such networks that is given 
in terms of a certain partially ordered set. 
This allows us to define two new spaces of
ranked binary tree-child networks. The first space can be considered as a
generalization of the recently introduced space of 
ranked binary phylogenetic trees whose distance is defined in terms of
ranked nearest neighbor interchange moves. 
The second space is a continuous space that captures all
equidistant tree-child networks and generalizes
the space of ultrametric trees. In particular,
we show that this continuous space is a so-called CAT(0)-orthant
space which, for 
example, implies that the distance between two equidistant tree-child networks
can be efficiently computed. 
\end{abstract}

\maketitle

\section{Introduction}
\label{sec:intro}

Rooted phylogenetic networks are essentially directed acyclic graphs,
whose leaf sets correspond to a set of species. They
are commonly used to represent evolutionary histories in which
reticulate events have occurred due to processes such as hybridization and lateral gene transfer. 
Various classes of rooted phylogenetic networks have been
defined, including the extensively studied class
of so-called {\em tree child networks} introduced by \citet{CRV08}  
(see e.g. \citet{kong2022classes} for a review). Recently, 
the class of (binary) {\em ranked tree child networks (RTCNs)}
was introduced by~\citet{BLS20a},
which have been further studied by \citet{CFY22}
and \citet{FLY2024}.
As their name suggests, these are a special type
of tree-child network that are endowed with additional information
which allows the evolutionary events represented by the 
network to be arranged consistently along a time line.
RTCNs generalize {\em ranked phylogenetic trees} 
(also called {\em ranked genealogies}), 
structures that can be  used to study evolutionary 
dynamics (see e.g. \citet{kim2020distance} and the references therein). 

\begin{figure}[h!]
\centering
\includegraphics[scale=1.0]{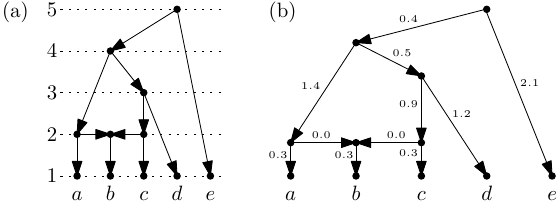}
\caption{(a) A binary RTCN on the set \(X = \{a,b,c,d,e\}\). Each
dotted horizontal line corresponds to vertices that have the same rank.
(b) An ETCN on~\(X\) obtained by assigning suitable weights to the
arcs of the RTCN in~(a).}
\label{fig:ex:intro:ranked}
\end{figure}

Informally (see Section~\ref{sec:rtcn} and Section~\ref{sec:consruction:network:space} for full definitions),
a binary RTCN is a binary rooted phylogenetic network with leaf set~$X$ 
having the following additional
restrictions: (i) every vertex that is not a leaf must be
the tail of some arc whose head has no other in-coming arcs and
(ii) vertices are assigned ranks from the set \(\{1,\dots,|X|\}\)
such that the tail of an arc never has a smaller rank than 
the head. In Figure~\ref{fig:ex:intro:ranked}(a) we give an example
of a binary RTCN. In addition, by assigning non-negative weights
to the arcs of an RTCN that are consistent with the
ranks of the vertices (in particular, vertices having the
same rank also have the same distance from the root)
we obtain an \emph{equidistant}
tree-child network (ETCN). An example of an ETCN
is given in Figure~\ref{fig:ex:intro:ranked}(b).
	
Comparing phylogenetic trees and networks
is an important problem in phylogenetics which has been studied for
some time, and various distances have been defined on trees and networks 
(see e.g. \citet{CRV08,HMW16a,JJEIS18a,kuhner2015practical,nakhleh2009metric,pons2019tree,smith2022robust}),
including ranked phylogenetic trees~\citep{kim2020distance}. 
Thus it is a natural question to ask for ways to compare RTCNs and ETCNs. 
In this paper, we shall present some new distances for 
such networks and consider some properties of the resulting spaces.
We define our distances by introducing a way to encode binary RTCNs 
on a fixed leaf set~\(X\) in terms of a certain partially ordered
set (or poset). As we shall see in Section~\ref{sec:encoding:binary:rtcn},
as well as encoding binary RTCNs, this poset has some
attractive mathematical properties, including the fact that it
generalizes the well-known poset
of partitions of the set~\(X\), a poset 
that captures the set of all binary ranked trees with leaf set~\(X\)
(see e.g.~\citet[Sec.~4.2]{huber2024space}).

Using our new encoding, in Section~\ref{sec:nni:on:rtcn}
we provide a generalization of the Robinson-Foulds distance on
rooted phylogenetic trees, and also define a generalization
of the ranked \emph{nearest neighbor interchange (rNNI)} distance on 
binary ranked trees introduced by~\citet{GWM18} to all binary RTCNs,
thus providing a way to compare binary RTCNs.
In addition, in Section~\ref{sec:consruction:network:space}
we define a continuous metric space
of ETCNs whose definition relies on some special properties
of the poset mentioned above.
More specifically, we show that this space
is a so-called \emph{CAT(0)-orthant space},
which implies that the distance between any two ETCNs 
can be computed efficiently. Note that \citet{BHV01}
presented a similar approach
to compare unrooted edge-weighted phylogenetic trees, and  
that our space of ETCNs generalizes the more recently
introduced spaces of ultrametric trees~\citep{GD16} and
equidistant cactuses~\citep{huber2024space}.

We now describe the contents of the rest of this paper.
After formally defining binary RTCNs in Section~\ref{sec:rtcn},
we show how binary RTCNs with a fixed leaf set correspond to
maximal chains of certain cluster systems
(Section~\ref{sec:encoding:binary:rtcn}). 
Then we introduce the poset capturing all binary RTCNs,
present our generalization of rNNIs and show
that the discrete space of all binary RTCNs is connected
under these more general rNNIs (Section~\ref{sec:nni:on:rtcn}).
Next we describe how our poset also systematically
captures certain non-binary RTCNs (Section~\ref{sec:non:binary:rtcn})
and use this to describe our CAT(0)-orthant space
of ETCNs (Section~\ref{sec:consruction:network:space}).
We conclude mentioning some possible directions for
future work (Section~\ref{sec:conclusion}).

\section{Binary ranked tree-child networks}
\label{sec:rtcn}

In this section, we formally define the basic type
of phylogenetic network that we consider in this paper. 
For the rest of this paper, \(X\)~will be a finite
non-empty set with \(n = |X| \geq 2\), which can be
thought of as a set of species or taxa.
	
A directed graph \(G=(V,E)\) consists of a finite, non-empty set~\(V\)
of \emph{vertices} and a set \(E \subseteq V \times V\) of
directed edges or \emph{arcs}. We write \((u,v)\) for an arc
that is directed from vertex~\(u\), the \emph{tail} of the arc, 
to vertex~\(v\), the \emph{head} of the arc.
For a vertex~\(u\), the \emph{out-degree} of~\(u\)
is the number of arcs that have~\(u\) as its tail and
the \emph{in-degree} of~\(u\) is the number of arcs that have~\(u\)
as its head. A \emph{leaf} is a vertex of out-degree~0. 
A \emph{directed path} in~\(G\) from vertex~\(s\) to
vertex~\(t\) is a sequence
\(s=v_1,v_2,\dots,v_k=t\) of \(k \geq 1\) pairwise distinct vertices with
\((v_i,v_{i+1}) \in E\) for all \(1 \leq i \leq k-1\). Note that we
allow \(k=1\), which then implies that \(s=t\).
A directed graph~\(G\) is \emph{acyclic} if it does not contain
a directed path from some vertex \(s\) to some vertex \(t\)
such that~\((t,s)\) is an arc in~\(G\) (which would then
form a directed cycle in~\(G\)). 

A \emph{rooted phylogenetic network}~\(\mathcal{N} = (V,E,\rho)\) on~\(X\)
is a directed acyclic graph \(G=(V,E)\) with leaf set~\(X\) and a unique
vertex~\(\rho\) of in-degree~0, called the \emph{root} of~\(\mathcal{N}\). 
A vertex of~\(\mathcal{N}\) that is not
a leaf is called an \emph{interior} vertex.
A vertex of~\(\mathcal{N}\) with in-degree at least 2 is a
\emph{hybrid vertex}. Any vertex of~\(\mathcal{N}\) that is
not a hybrid vertex is a \emph{tree vertex}.
A rooted phylogenetic network is \emph{binary} if 
the root has out-degree~2 and every other interior vertex
either has in-degree~1 and out-degree~2 or in-degree~2 and out-degree~1.

\begin{figure}
\centering
\includegraphics[scale=1.0]{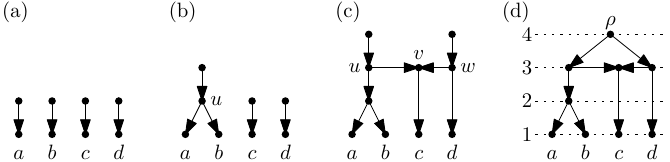}
\caption{An example of the process that generates a binary RTCN
on \(X=\{a,b,c,d\}\).
(a) The result of Step~1.
(b) The result of performing~(1) in Step~2.
(c) The result of performing~(2) in Step~3.
(d) The resulting binary RTCN after Step~\(n=4\). Vertices
of rank~\(i\) are drawn on the dotted horizontal line
numbered~\(i\) \((1 \leq i \leq 4)\).}
\label{fig:def:rtcn}
\end{figure}

We now define what we mean by a
binary \emph{ranked tree-child network} (RTCN)
following \citep[Sec. 2.1]{BLS20a}, which describes
how any such network on a fixed set~\(X\) can be obtained using
a  process involving \(n\)~steps:
\begin{itemize}
\item
\underline{Step 1:}
For each \(x \in X\) an arc with head~\(x\) is created. The tails
of these arcs are pairwise distinct and form a
set of \(n\)~vertices with in-degree~0 (see Figure~\ref{fig:def:rtcn}(a)).
\item
\underline{Step $i$ $(2 \leq i \leq n-1)$:}
Precisely one of the following modifications to the network obtained
in Step~\(i-1\) is performed:
\begin{itemize}
\item[(1)]
Two vertices with in-degree~0 are selected.
These two vertices are identified as a single
vertex~\(u\) with out-degree~2. Then a new arc
with head~\(u\) and a new vertex as its tail is added
(see Figure~\ref{fig:def:rtcn}(b)).
\item[(2)]
Three vertices \(u\), \(v\) and \(w\) with in-degree~0 are selected.
Then arcs \((u,v)\) and \((w,v)\) are added, making~\(v\) a
hybrid vertex. Then two new arcs with head~\(u\) and~\(w\),
respectively, and each with a new vertex as its tail are
added (see Figure~\ref{fig:def:rtcn}(c)).
\end{itemize}
After performing Step~\(i\) we have a network that
has \(n-i+1\) vertices with in-degree~0.
\item
\underline{Step $n$:}
The result of Step~\(n-1\) is a network with precisely two
vertices with in-degree~0. These two vertices are identified
as a single vertex which then forms the root~\(\rho\) of
the resulting binary RTCN. This finishes the process of
generating a binary RTCN (see Figure~\ref{fig:def:rtcn}(d)). 
\end{itemize}

All different binary RTCNs on a fixed set~\(X\) arise
through the choice of performing either~(1) or~(2) in Steps~\(2,\dots,n-1\)
and, subsequently, the choice of either the two vertices used in~(1) or
the three vertices used on~(2). Note that in~(2)
the role of vertex~\(v\) is different from the roles
of vertices~\(u\) and~\(w\). So, more precisely,
performing~(2) also involves a choice which of the three selected
vertices plays the role of the vertex that becomes
a hybrid vertex. If~(2) is never performed in any of
the Steps~\(2,\dots,n-2\) the network only contains
tree vertices and is called a \emph{binary ranked tree}.

Each vertex~\(q\) in a binary
RTCN~\(\mathcal{N} = (V,E,\rho)\) on~\(X\) has a \emph{rank} from
the set \(\{1,\dots,n\}\) associated with it that is
denoted by \(\text{rank}(q)\). More precisely
(see Figure~\ref{fig:def:rtcn}(d)), we have
\begin{itemize}
\item
\(\text{rank}(x)=1\) for all \(x \in X\),
\item
\(\text{rank}(u)=i\) when~(1) is performed in Step~\(i\)
\((2 \leq i \leq n-1)\),
\item
\(\text{rank}(u)=\text{rank}(v)=\text{rank}(w)=i\)
when~(2) is performed in Step~\(i\) \((2 \leq i \leq n-1)\), and
\item
\(\text{rank}(\rho) = n\).
\end{itemize}
These ranks correspond to an ordering of the biological
events (speciation or hybridization) that led from the
common ancestor at the root of the network to the elements in~\(X\) at
the leaves. The term \emph{tree-child} refers to the
fact that in the networks generated by the process
described above every interior vertex is the tail
of an arc whose head is a tree-vertex.
Note that tree-child networks without ranked vertices
were introduced by \citet{CRV08} and
remain an active area of research
(see e.g. \citet{CPS19,CZ20,FYZ21}).

\section{Encoding binary ranked tree-child networks}
\label{sec:encoding:binary:rtcn}

In this section, we present a way to \emph{encode} binary RTCNs, that is, 
a way to describe binary RTCNs in such a way that two RTCNs are the same if and
only if they have the same description. The encoding itself
is a straight-forward translation of the process described
in Section~\ref{sec:rtcn} for generating a binary RTCN
into the language of collections of subsets of~\(X\). 
As we will see later on, this encoding
is very helpful for proving our results about RTCNs.

To formally describe the encoding, we first give some more
definitions. A \emph{cluster} on \(X\) is a non-empty subset of \(X\).
A \emph{cluster system} on~\(X\)
is a non-empty collection of clusters on~\(X\). Given a 
rooted phylogenetic network
\(\mathcal{N}=(V,E,\rho)\) on~\(X\),
to each each vertex~\(v \in V\), we associate 
the cluster~\(C_{v}\) on~\(X\) that consists of all those \(x \in X\) for which
there exists a directed path in~\(\mathcal{N}\) from~\(v\) to~\(x\).
The clusters given in this way are sometimes called
the {\em hard-wired} clusters
of the network. 

Each step \(i\) \((1 \leq i \leq n)\)
in the process described in Section~\ref{sec:rtcn}
can now be captured by a cluster system 
\(\mathcal{C}_i\) on~\(X\) as follows:
\begin{itemize}
\item
\underline{Step 1:}
\(\mathcal{C}_1 = \{\{x\} : x \in X\}\). Each cluster in
\(\mathcal{C}_1\) consists of a single element and
represents a leaf in the resulting network.
\item
\underline{Step $i$ $(2 \leq i \leq n-1)$:}
We already have the cluster system \(\mathcal{C}_{i-1}\) which
consists of the clusters~\(C_v\) obtained from those
vertices~\(v\) that are the head of an arc whose tail has
in-degree~0 at the end of Step~\(i-1\). 
\begin{itemize}
\item
If~(1) is performed in Step~\(i\) there must exist
clusters \(A\) and \(B\) in \(\mathcal{C}_{i-1}\) such that
\(C_u = A \cup B\). Then we put
\(\mathcal{C}_i = (\mathcal{C}_{i-1} - \{A,B\}) \cup \{C_u\}\).
\item
If~(2) is performed in Step~\(i\) there must exist
clusters \(A\), \(B\) and \(C\) in \(\mathcal{C}_{i-1}\) such that
\(C_u = A \cup B\) and \(C_w = B \cup C\). Then we put
\(\mathcal{C}_i = (\mathcal{C}_{i-1} - \{A,B,C\}) \cup \{C_u,C_w\}\).
\end{itemize}
The cluster system \(\mathcal{C}_i\) consists of \(n-i+1\) clusters.
\item
\underline{Step $n$:}
The cluster system \(\mathcal{C}_{n-1}\) consists of two clusters
\(A\) and \(B\) such that \(C_{\rho} = A \cup B = X\). We put
\(\mathcal{C}_n = (\mathcal{C}_{n-1} - \{A,B\}) \cup \{C_{\rho}\} = \{X\}\).
\end{itemize}
To illustrate this definition, consider again the example of generating
a binary RTCN on \(X=\{a,b,c,d\}\) in Figure~\ref{fig:def:rtcn}. Then
we obtain the following cluster systems:
\begin{align*}
\mathcal{C}_1 &= \{\{a\},\{b\},\{c\},\{d\}\},\\
\mathcal{C}_2 &= (\mathcal{C}_1 - \{\{a\},\{b\}\}) \cup \{\{a,b\}\} = \{\{a,b\},\{c\},\{d\}\},\\
\mathcal{C}_3 &= (\mathcal{C}_2 - \{\{a,b\},\{c\},\{d\}\}) \cup \{\{a,b,c\},\{c,d\}\} = \{\{a,b,c\},\{c,d\}\}, \mbox{ and }\\
\mathcal{C}_4 &= (\mathcal{C}_3 - \{\{a,b,c\},\{c,d\}\}) \cup \{\{a,b,c,d\}\} =\{\{a,b,c,d\}.\} 
\end{align*}
As also illustrated by this example,
the cluster systems can easily be read from the resulting binary RTCN
\(\mathcal{N} = (V,E,\rho)\) on~\(X\): For \(1 \leq i < n\), 
we have
\[\mathcal{C}_i = \mathcal{C}_i(\mathcal{N}) = 
\{C_{v} : \text{there exists an arc} \ (u,v)
\in E \ \text{with rank}(u) > i \geq \text{rank}(v)\},\]
and \(\mathcal{C}_n = \{C_{\rho}\} = \{X\}\).

\begin{figure}
\centering
\includegraphics[scale=1.0]{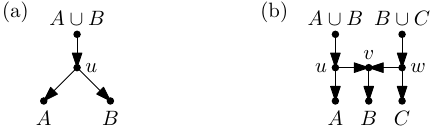}
\caption{The two operations (a) \(\vdash_{(1)}\)
and (b) \(\vdash_{(2)}\) that can be applied to a cluster system
and how they are related to the process of generating binary RTCNs.}
\label{fig:coal:op:types}
\end{figure}

To make more precise our way of encoding a binary RTCN~\(\mathcal{N}\) by
the cluster systems
\[\mathcal{C}_1(\mathcal{N}),\dots,\mathcal{C}_n(\mathcal{N}),\]
we need a little bit more notation.
Let \(\mathcal{C}\) and \(\mathcal{C}'\) be cluster systems on~\(X\).
We write:
\begin{itemize}
\item
\(\mathcal{C} \vdash_{(1)} \mathcal{C}'\) if
there exist two distinct clusters \(A,B \in \mathcal{C}\) with
\(\mathcal{C}' = (\mathcal{C} - \{A,B\}) \cup \{A \cup B\}\)
(see Figure~\ref{fig:coal:op:types}(a)).
\item
\(\mathcal{C} \vdash_{(2)} \mathcal{C}'\) if
there exist three pairwise distinct clusters \(A,B,C \in \mathcal{C}\) with
\(\mathcal{C}' = (\mathcal{C} - \{A,B,C\}) \cup \{A \cup B,B \cup C\}\)
(see Figure~\ref{fig:coal:op:types}(b)).
\end{itemize}
We will often use the simplified notation
\(\mathcal{C} \vdash \mathcal{C}'\) if
either \(\mathcal{C} \vdash_{(1)} \mathcal{C}'\) or
\(\mathcal{C} \vdash_{(2)} \mathcal{C}'\) holds 
in case it is not relevant which of the two conditions holds. 
A \emph{maximal chain} on~\(X\) is a
sequence \(\mathcal{C}_1,\dots,\mathcal{C}_n\)
of~\(n\) cluster systems on~\(X\) such that
\[\{\{x\} : x \in X\} = \mathcal{C}_1 \vdash \mathcal{C}_2 \vdash \dots
\vdash \mathcal{C}_n = \{X\}.\]

Before we state the main result of this section, we establish
a useful property of cluster systems in a maximal chain on~\(X\).

\begin{lemma}
\label{lem:unique:elements}
Let \(\mathcal{C}_1,\dots,\mathcal{C}_n\) be a maximal chain on~\(X\)
and \(1 \leq i \leq n\). Then every cluster in~\(\mathcal{C}_i\)
contains an element in $X$ that is not contained
in any other cluster in~\(\mathcal{C}_i\).
\end{lemma}

\pf
We use induction on~\(i\). In the base case of
the induction, \(i=1\), we have \(\mathcal{C}_1=\{\{x\} : x \in X\}\).
Then, clearly, for all \(x \in X\), the element \(x\)
is only contained in the cluster~\(\{x\}\).

Next consider the case \(i>1\).
By the definition of a maximal chain,
we have \(\mathcal{C}_{i-1} \vdash \mathcal{C}_i\).
By induction, all clusters in \(\mathcal{C}_{i-1}\) contain at least one
element that is not contained in any other cluster in~\(\mathcal{C}_{i-1}\).
But then, in view of the definition of \(\vdash_{(1)}\) and \(\vdash_{(2)}\),
it follows that also all clusters in \(\mathcal{C}_i\) contain at least one
element that is not contained in any other cluster in~\(\mathcal{C}_i\),
as required.
\epf

We now present our encoding for binary RTCNs. 

\begin{theorem}
\label{thm:chain:encoding}
Binary RTCNs on~\(X\) are in bijective correspondence with
maximal chains on~\(X\).
\end{theorem}

\pf
We have already seen that from every binary RTCN \(\mathcal{N}\) on~\(X\) we
obtain the maximal chain \(\mathcal{C}_1(\mathcal{N}),\dots,\mathcal{C}_n(\mathcal{N})\) on~\(X\).

So, assume that \(\mathcal{C}_1,\dots,\mathcal{C}_n\) is a 
maximal chain on~\(X\).
To obtain a binary RTCN~\(\mathcal{N}\) on~\(X\) with
\(\mathcal{C}_i = \mathcal{C}_i(\mathcal{N})\) we use the
maximal chain to
guide the process of generating~\(\mathcal{N}\) during
Steps \(i=2,\dots,n-1\):
\begin{itemize}
\item
If \(\mathcal{C}_{i-1} \vdash_{(1)} \mathcal{C}_i\) we perform~(1).
\item
If \(\mathcal{C}_{i-1} \vdash_{(2)} \mathcal{C}_i\) we perform~(2).
\end{itemize}
It remains to show that the two vertices with in-degree~0
used when performing~(1) and the three vertices with
in-degree~0 used when performing~(2), respectively, are
uniquely determined by the maximal chain on~\(X\). 
But this follows immediately from the property of the
cluster systems in a maximal chain on~\(X\) stated in
Lemma~\ref{lem:unique:elements}, as this allows to
uniquely determine the clusters involved in
\(\mathcal{C}_{i-1} \vdash \mathcal{C}_i\).
\epf

The encoding established in Theorem~\ref{thm:chain:encoding}
is useful because it allows us to systematically break
any binary RTCN on~\(X\) down into building blocks (i.e. 
cluster systems), which gives a simple way to understand the
relationship between two binary RTCNs. 
We remark that there are two interesting special instances of our encoding:
\begin{itemize}
\item
Maximal chains \(\mathcal{C}_1,\dots,\mathcal{C}_n\) on~\(X\) such that
\(\mathcal{C}_1 \vdash_{(1)} \mathcal{C}_2 \vdash_{(1)} \dots \vdash_{(1)}
\mathcal{C}_n\) are in bijective correspondence with binary
ranked trees on~\(X\).
\item
Let~\(\vdash^*\) be the restricted variant of~\(\vdash\)
defined by the additional requirement that
\begin{itemize}
\item
for \(\mathcal{C} \vdash^*_{(1)} \mathcal{C}'\) to hold we must have
\(A \cap B \neq \emptyset\) or \(A \cap C = \emptyset\) for
all \(C \in \mathcal{C} - \{A\}\) or \(B \cap C = \emptyset\) for
all \(C \in \mathcal{C} - \{B\}\), and 
\item
for \(\mathcal{C} \vdash^*_{(2)} \mathcal{C}'\) to hold we must have
\(A \cap D = \emptyset\) for all \(D \in \mathcal{C} - \{A\}\),
\(B \cap D = \emptyset\) for all \(D \in \mathcal{C} - \{B\}\) and
\(C \cap D = \emptyset\) for all \(D \in \mathcal{C} - \{C\}\).
\end{itemize}
Then maximal chains \(\mathcal{C}_1,\dots,\mathcal{C}_n\) on~\(X\) such that
\(\mathcal{C}_1 \vdash^* \mathcal{C}_2 \vdash^* \dots \vdash^*
\mathcal{C}_n\) are in bijective correspondence with 
\emph{binary ranked cactuses} on~\(X\), a proper subclass of binary
RTCNs considered by~\citet{huber2024space}.
\end{itemize}

\section{Nearest neighbor interchange moves for binary RTCNs}
\label{sec:nni:on:rtcn}

In this section we explain how to use our
encoding of binary RTCNs by maximal chains of cluster systems to compare 
unweighted binary RTCNs. 
One simple way to do this is to define the distance 
between two such networks $\mathcal N$ and $\mathcal N'$ to be
\[
|\{\mathcal{C}_1(\mathcal{N}), \dots, \mathcal{C}_n(\mathcal{N})\}
\triangle \{\mathcal{C}_1(\mathcal{N}'), \dots, \mathcal{C}_n(\mathcal{N}')\}|
\]
where $\triangle$ denotes the symmetric difference of sets.
The metric on binary RTCNs arising in this way can be
thought of as a ranked analogue 
of the Robinson-Foulds distance on
rooted trees \citep{robinson1981comparison}.

\begin{figure}[b]
\centering
\includegraphics[scale=1.0]{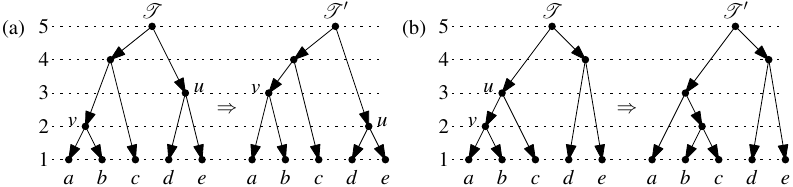}
\caption{The two types of modifications on binary ranked trees
allowed in an rNNI:
(a) Swapping the ranks of vertices~\(u\) and~\(v\).
(b) An actual nearest neighbor interchange.}
\label{fig:nni:trees}
\end{figure}

A more sophisticated approach is to define an analogue
of the well-known nearest neighbor interchange distance for
rooted phylogenetic trees \citep{robinson1971comparison}.
This distance has already been 
generalized to binary ranked trees by~\citet{GWM18} as follows.
First, define two types of modifications of a binary ranked tree on~\(X\)
(called \emph{ranked nearest neighbor interchanges (rNNIs)}): 
\begin{itemize}
\item
For two vertices \(u\) and \(v\) with
\(\text{rank}(u) = \text{rank}(v) + 1\) and
\((u,v)\) not an arc, the ranks of \(u\) and
\(v\) are swapped without changing the
topology of the tree (see Figure~\ref{fig:nni:trees}(a)).
\item
For two vertices \(u\) and \(v\) with
\(\text{rank}(u) = \text{rank}(v) + 1\) and
\((u,v)\) an arc, the topology of the tree 
is changed (see Figure~\ref{fig:nni:trees}(b)).
\end{itemize}

Then, \citet{GWM18} established the following result:

\begin{fact}
\label{fact:rnni:connected:trees}
For any two binary ranked trees~\(\mathcal{T}\)
and~\(\mathcal{T}'\) on~\(X\) there exists
a sequence of rNNIs that transform~\(\mathcal{T}\) into~\(\mathcal{T}'\).
\end{fact}

Interestingly, as pointed out in the
supplementary material by~\citet{CEFB21a},
there is a concise and uniform way to describe
an rNNI between two binary ranked trees \(\mathcal{T}\)
and \(\mathcal{T}'\) on~\(X\) using the corresponding maximal chains
\(\mathcal{C}_1,\dots,\mathcal{C}_n\) and
\(\mathcal{C}'_1,\dots,\mathcal{C}'_n\) on~\(X\): There
exists \(2 \leq i \leq n-1\) such that
\(\mathcal{C}_i \neq \mathcal{C}'_i\) and
\(\mathcal{C}_j = \mathcal{C}'_j\) for all \(j \neq i\).
Less formally, there is an rNNI between \(\mathcal{T}\)
and \(\mathcal{T}'\) if the corresponding maximal chains differ
in precisely one cluster system. For example, 
consider the two binary ranked trees
\(\mathcal{T}\) and \(\mathcal{T}'\)
on~\(X=\{a,b,c,d,e\}\) in Figure~\ref{fig:nni:trees}(a).
Looking at the corresponding maximal chains on~\(X\) we have:
\begin{align*}
\mathcal{C}_1 &= \{\{a\},\{b\},\{c\},\{d\},\{e\}\} = \mathcal{C}'_1\\
\mathcal{C}_2 &= \{\{a,b\},\{c\},\{d\},\{e\}\} \neq \{\{a\},\{b\},\{c\},\{d,e\}\} = \mathcal{C}'_2\\
\mathcal{C}_3 &= \{\{a,b\},\{c\},\{d,e\}\} = \mathcal{C}'_3\\
\mathcal{C}_4 &= \{\{a,b,c\},\{d,e\}\} = \mathcal{C}'_4\\
\mathcal{C}_5 &= \{\{a,b,c,d,e\}\} = \mathcal{C}'_5
\end{align*}

\begin{figure}[b]
\centering
\includegraphics[scale=1.0]{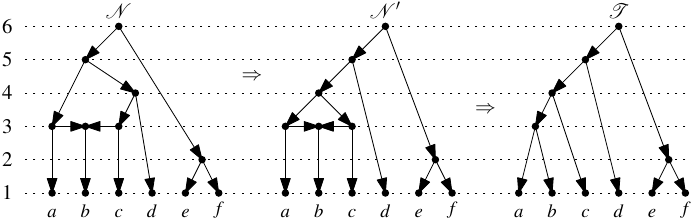}
\caption{Two consecutive rNNI$^*$s that transform the binary
RTCN~\(\mathcal{N}\) on \(X=\{a,b,c,d,e,f\}\) through the
intermediate binary RTCN~\(\mathcal{N}'\) to
the binary ranked tree~\(\mathcal{T}\).}
\label{fig:ex:rnni}
\end{figure}

While the description of rNNIs in terms of the binary
ranked trees is very intuitive, it is not obvious how
to directly generalize this to binary RTCNs. 
However, as with ranked trees, 
the description in terms of maximal chains on~\(X\) immediately
suggests a way to do this: We say that
there is a \emph{ranked nearest neighbor interchange}
between two binary RTCNs \(\mathcal{N}\) and \(\mathcal{N}'\) (both on \(X\))
if the corresponding maximal chains on~\(X\)
differ in precisely one cluster system.
We shall continue to use rNNI when referring to
ranked nearest neighbor interchanges
restricted to binary ranked trees as described above
and will use rNNI$^*$ when referring to this generalization. 
Figure~\ref{fig:ex:rnni} illustrates what happens to the
corresponding RTCNs when we apply such rNNI$^*$s.

We conclude this section by establishing that
for any two binary RTCNs on~\(X\) there exists
a sequence of rNNI$^*$s that transforms one into the other.
This implies that we can define a distance between any pair of binary RTCNs
by taking the length of a shortest sequence of rNNI$^*$s 
that transforms one of the networks to the other.

\begin{theorem}
\label{thm:gallery:connected}
For any two binary RTCNs~\(\mathcal{N}_1\) and~\(\mathcal{N}_2\)
on~\(X\) there exists
a sequence of rNNI$^*$s that transform~\(\mathcal{N}_1\) into~\(\mathcal{N}_2\).
\end{theorem} 

\pf
Since every rNNI between two binary ranked trees on~\(X\)
is also an rNNI$^*$ between them when we view the binary ranked trees
as binary RTCNs, it suffices, by Fact~\ref{fact:rnni:connected:trees},
to show that for any binary RTCN \(\mathcal{N}\) on~\(X\)
there exists a sequence of rNNI$^*$s that transforms~\(\mathcal{N}\)
into some binary ranked tree~\(\mathcal{T}\) on~\(X\)
(Figure~\ref{fig:ex:rnni} gives an example of such a sequence
of rNNI$^*$s).

Let \(\mathcal{C}_1, \dots, \mathcal{C}_n\) 
be the maximal chain on~\(X\) that corresponds to \(\mathcal{N}\)
by Theorem~\ref{thm:chain:encoding}.
The proof is by induction on the number \(\ell\) of those
\(1 < j \leq n\) with \(\mathcal{C}_{j-1} \vdash_{(2)} \mathcal{C}_{j}\).
In the base case of the induction, \(\ell=0\), 
\(\mathcal{N}\) is itself a binary ranked tree.

So, assume that \(\ell > 0\).
Let~\(i\) be the maximum of those \(1 < j \leq n\) 
with \(\mathcal{C}_{j-1} \vdash_{(2)} \mathcal{C}_{j}\).
By the definition of \(\vdash_{(2)}\) there exist three
pairwise distinct \(A,B,C \in \mathcal{C}_{i-1}\) such that
\[\mathcal{C}_{i} = (\mathcal{C}_{i-1}-\{A,B,C\}) \cup \{A \cup B,B \cup C\}.\]
By Lemma~\ref{lem:unique:elements}, we can select from each
cluster in \(\mathcal{C}_i\) an element that is unique to this
cluster. Let~\(X' \subseteq X\) be the resulting subset
of selected elements. To give an example, for the binary RTCN~\(\mathcal{N}\)
in Figure~\ref{fig:ex:rnni} we have \(i=3\) and can select~\(X'=\{a,c,d,f\}\).

\begin{figure}
\centering
\includegraphics[scale=1.0]{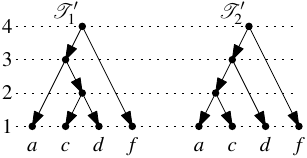}
\caption{The ranked trees \(\mathcal{T}_1'\) and
\(\mathcal{T}_2'\) referred to in the proof of
Theorem~\ref{thm:gallery:connected}.}
\label{fig:restriction:rnni}
\end{figure}

By the maximality of~\(i\), we have
\(\mathcal{C}_{j-1} \vdash_{(1)} \mathcal{C}_{j}\) for
all \(j > i\). Thus, after restricting all clusters
to~\(X'\), the sequence \(\mathcal{C}_{i},\dots,\mathcal{C}_{n}\)
becomes a maximal chain on~\(X'\) that only contains
partitions of~\(X'\). This maximal chain on~\(X'\)
corresponds, by Theorem~\ref{thm:chain:encoding}, to a
binary ranked tree~\(\mathcal{T}_1'\) on~\(X'\).
In Figure~\ref{fig:restriction:rnni} the binary ranked
tree~\(\mathcal{T}_1'\) resulting from the binary RTCN~\(\mathcal{N}\)
in Figure~\ref{fig:ex:rnni} is shown.

Let \(y\) be the element in \(X'\) selected from \(A \cup B\) and
let \(z\) be the element in \(X'\) selected from \(B \cup C\).
Let \(\mathcal{T}_2'\) be a binary ranked tree on~\(X'\)
that contains a vertex~\(u\) with \(\text{rank}(u)=2\) and the
arcs \((u,y)\) and \((u,z)\). Clearly, such a binary ranked
tree exists and, by Fact~\ref{fact:rnni:connected:trees},
there exists a sequence of rNNIs that transforms \(\mathcal{T}_1'\)
into \(\mathcal{T}_2'\). In Figure~\ref{fig:restriction:rnni},
a suitable binary ranked tree~\(\mathcal{T}_2'\) is shown
that arises by applying a single rNNI to~\(\mathcal{T}_1'\).

The sequence of rNNIs transforming~\(\mathcal{T}_1'\)
into~\(\mathcal{T}_2'\) corresponds to a sequence
of rNNI$^*$s that transform~\(\mathcal{N}\) into a binary
RTCN \(\mathcal{N}'\) such that all vertices of~\(\mathcal{N}\)
with rank at most~\(i\) remain unchanged and only the 
vertices corresponding to the binary ranked tree~\(\mathcal{T}_1'\)
are involved. In Figure~\ref{fig:ex:rnni} the binary
RTCN~\(\mathcal{N}'\) resulting from the rNNI between the
binary ranked trees~\(\mathcal{T}_1'\) and~\(\mathcal{T}_2'\)
in Figure~\ref{fig:restriction:rnni} is shown. 

In preparation for the last step in the proof, we summarize the properties
of the maximal chain on~\(X\) that corresponds to~\(\mathcal{N}'\):
\begin{itemize}
\item
\(\mathcal{C}_j(\mathcal{N}') = \mathcal{C}_j(\mathcal{N})\)
for all \(1 \leq j \leq i\)
\item
\(\mathcal{C}_{j-1}(\mathcal{N}') \vdash_{(1)} \mathcal{C}_j(\mathcal{N}')\)
for all \(i < j \leq n\)
\item
\(\mathcal{C}_{i+1}(\mathcal{N}')
= (\mathcal{C}_{i}(\mathcal{N}') - \{A \cup B,B \cup C\}) \cup \{A \cup B \cup C\}\)
\end{itemize}

Now we perform the following rNNI$^*$ on~\(\mathcal{N}'\):
We replace the cluster system
\(\mathcal{C}_i = \mathcal{C}_{i}(\mathcal{N}')\)
by the cluster system
\[\mathcal{C}_i'' = (\mathcal{C}_{i-1}(\mathcal{N}') - \{A,B\}) \cup \{A \cup B\}.\]
This is possible since \(A,B,C \in \mathcal{C}_{i-1}(\mathcal{N}')\).
Then we have
\[\mathcal{C}_{i+1}(\mathcal{N}') = (\mathcal{C}_i'' - \{A \cup B,C\}) \cup \{A \cup B \cup C\}.\]
The resulting maximal chain on~\(X\) is
\[\mathcal{C}_1(\mathcal{N}') \vdash \dots \vdash \mathcal{C}_{i-1}(\mathcal{N}')
\vdash_{(1)} \mathcal{C}_i'' \vdash_{(1)} \mathcal{C}_{i+1}(\mathcal{N}') \vdash_{(1)} \dots \vdash_{(1)} \mathcal{C}_{n}(\mathcal{N}').\]
By Theorem~\ref{thm:chain:encoding}, this maximal chain on~\(X\)
corresponds to a binary RTCN~\(\mathcal{N}''\)
on~\(X\). Moreover, by construction,
the number of occurrences of \(\vdash_{(2)}\) in this maximal chain on~\(X\)
is \(\ell-1\). Hence, by induction, there exists a sequence of
rNNI$^*$s that transform~\(\mathcal{N}''\) into a binary ranked
tree~\(\mathcal{T}\) on~\(X\). But then, there is also a 
sequence of rNNI$^*$s that transform~\(\mathcal{N}\) into~\(\mathcal{T}\).
This finishes the inductive proof. In the example
in Figure~\ref{fig:ex:rnni} we have \(\mathcal{N}'' = \mathcal{T}\).
\epf

\section{Non-binary ranked tree-child networks}
\label{sec:non:binary:rtcn}

So far we have shown how to compare binary RTCNs whose arcs
are unweighted. In order to compare ETCNs, we will
need to consider non-binary RTCNs, as these can arise when
shrinking down edges to length zero.
In general, non-binary rooted phylogenetic networks
tend to be harder to capture than their binary counter-parts
(see e.g. \citet{jetten2016nonbinary}). 
In this section, we define certain partially ordered set or
\emph{poset} which not only allows us to handle non-binary
RTCNs, but also to define a distance on ETCNs in the next section.

First, suppose that~\(\mathfrak{T}(X)\) denotes the set of all cluster systems on~\(X\) that
occur in some maximal chain on~\(X\).
For \(\mathcal{C},\mathcal{C}' \in \mathfrak{T}(X)\)
we write \(\mathcal{C} \preceq \mathcal{C}'\) if there exists
a maximal chain
\[\mathcal{C}_1 \vdash \mathcal{C}_2 \vdash \dots \vdash \mathcal{C}_n\]
on~\(X\) with \(\mathcal{C} = \mathcal{C}_i\) and
\(\mathcal{C}'= \mathcal{C}_j\) for some \(1 \leq i \leq j \leq n\).
Then, by construction, \(\preceq\)~is
a partial ordering on~\(\mathfrak{T}(X)\). We
denote the resulting poset by \((\mathfrak{T}(X),\preceq)\).
Note that a \emph{chain}
in \((\mathfrak{T}(X),\preceq)\) is a sequence
\(\mathcal{C}_1,\dots,\mathcal{C}_t\) of \(2 \leq t \leq n\)
pairwise distinct cluster systems in \(\mathfrak{T}(X)\)
such that
\[\{\{x\}:x \in X\} = \mathcal{C}_1 \preceq \mathcal{C}_2 \preceq \dots \preceq \mathcal{C}_t = \{X\}.\]
The integer~\(t\) is called the \emph{length} of the chain.
Thus, chains of length~\(n\) in \((\mathfrak{T}(X),\preceq)\) are
precisely the maximal chains on~\(X\).

\begin{example}
\label{ex:non:maximal:chain}
Consider \(X=\{a,b,\dots,h\}\). Then
\begin{align*}
\mathcal{C}_1 &= \{\{a\},\{b\},\{c\},\{d\},\{e\},\{f\},\{g\},\{h\}\}\\
\mathcal{C}_2 &= \{\{a,b,c,d\},\{c,d,e\},\{f\},\{g,h\}\}\\
\mathcal{C}_3 &= \{\{a,b,c,d,e\},\{f,g,h\}\}\\
\mathcal{C}_4 &= \{\{a,b,c,d,e,f,g,h\}\}
\end{align*}
is a chain of length~4 in~\((\mathfrak{T}(X),\preceq)\).
\end{example}

In the proof of Theorem~\ref{thm:chain:encoding},
we saw how a maximal chain on~\(X\) guides the process
of generating the binary RTCN on~\(X\) that corresponds to
the maximal chain. Here we generalize this idea
to \emph{all} chains in~\((\mathfrak{T}(X),\preceq)\).
Since we may no longer
have \(\mathcal{C}_i \vdash \mathcal{C}_{i+1}\)
for two consecutive cluster systems in a chain, however,
the process of generating the RTCN corresponding to
a chain becomes a bit more complex to describe.

Let \(\mathcal{C}_1,\dots,\mathcal{C}_t\) be a chain
in~\((\mathfrak{T}(X),\preceq)\). The process of generating
the corresponding RTCN consists of~\(t\) steps:

\begin{figure}
\centering
\includegraphics[scale=1.0]{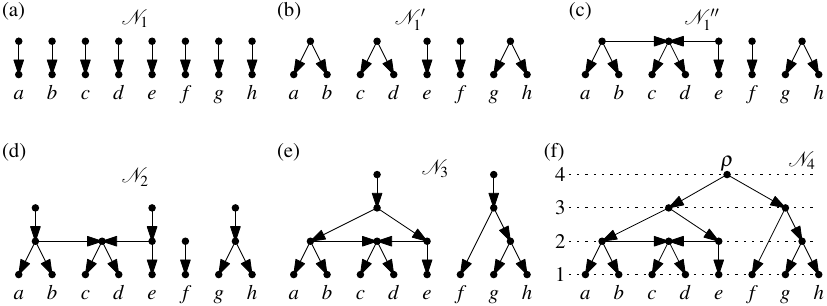}
\caption{The process that generates a non-binary
RTCN on~\(X=\{a,b,\dots,h\}\) from the chain of length~4
in Example~\ref{ex:non:maximal:chain}.
(a) The result of performing Step~1.
(b) The result of Phase~1 in Step~2.
(b) The result of Phase~2 in Step~2.
(d) The result of performing Step~2.
(e) The result of performing Step~3.
(f) The resulting non-binary RTCN after performing Step~4, the final step.}
\label{fig:process:non:binary:rtcn}
\end{figure}

\begin{itemize}
\item
\underline{Step 1:}
For each \(x \in X\) an arc with head~\(x\) is created. The tails
of these arcs are pairwise distinct and form a
set of \(n\)~vertices with in-degree~0
(see Figure~\ref{fig:process:non:binary:rtcn}(a)).
\item
\underline{Step $i$ $(2 \leq i \leq t-1)$:}
Let \(\mathcal{N}_{i-1}\) denote the network obtained
at the end of Step~$i-1$. For vertices~\(v\) of~\(\mathcal{N}_{i-1}\)
we also use~\(C_v\) to denote the
cluster on~\(X\) consisting of those \(x \in X\)
for which there exists a directed path in \(\mathcal{N}_{i-1}\)
from~\(v\) to~\(x\). There is a bijective correspondence
between the vertices~\(v\) with in-degree~0 of~\(\mathcal{N}_{i-1}\)
and the clusters in~\(\mathcal{C}_{i-1}\) obtained by mapping~\(v\)
to~\(C_v\). For all \(A \in \mathcal{C}_{i-1}\) put
\[H(A) = \{B \in \mathcal{C}_i : A \subseteq B\}.\]
It follows from the definition of \(\preceq\) 
that \(H(A) \neq \emptyset\) for all \(A \in \mathcal{C}_{i-1}\).
Moreover, by Lemma~\ref{lem:unique:elements},
for all \(B \in \mathcal{C}_i\) there exists some
\(A \in \mathcal{C}_{i-1}\) with \(H(A) = \{B\}\).
To illustrate the notation used to describe Step~\(i\),
consider~\(i=2\) for Example~\ref{ex:non:maximal:chain}
where we have:
\begin{align*}
\quad \quad \quad H(\{a\}) &= \{\{a,b,c,d\}\} = H(\{b\}), \
H(\{c\}) = \{\{a,b,c,d\},\{c,d,e\}\} = H(\{d\}),\\
\quad \quad \quad H(\{e\}) &= \{\{c,d,e\}\}, \ H(\{f\}) = \{\{f\}\}, \ H(\{g\})  = \{\{g,h\}\} = H(\{h\})
\end{align*}
Step~\(i\) consists of three phases:
\begin{itemize}
\item
\emph{Phase 1}: Any two vertices~\(v\) and~\(v'\) of~\(\mathcal{N}_{i-1}\)
with in-degree~0 are identified if \(H(C_v) = H(C_{v'})\).
Let~\(\mathcal{N}_{i-1}'\) denote the resulting network
(see Figure~\ref{fig:process:non:binary:rtcn}(b)).
For all vertices~\(u\) of~\(\mathcal{N}_{i-1}'\) with in-degree~0,
let \(H_u\) denote the set \(H(C_v)\), where~\(v\) is any of the vertices
of~\(\mathcal{N}_{i-1}\) with in-degree~0 that have been identified to
form~\(u\).
\item
\emph{Phase 2}: For all vertices~\(u\) of~\(\mathcal{N}_{i-1}'\)
with in-degree~0 and \(|H_u| \geq 2\) and for all
vertices~\(u'\) of~\(\mathcal{N}_{i-1}'\) with in-degree~0,
\(|H_{u'}| = 1\) and \(H_{u'} \subseteq H_u\), add the arc
with head~\(u\) and tail~\(u'\). Since \(|H_u| \geq 2\),
the vertices~\(u\) in this phase will become hybrid vertices.
Let~\(\mathcal{N}_{i-1}''\) denote the resulting network
(see Figure~\ref{fig:process:non:binary:rtcn}(c)).
\item
\emph{Phase 3}:
For all vertices~\(u\) of~\(\mathcal{N}_{i-1}''\)
with in-degree~0 and out-degree at least~2,
add a new arc with head~\(u\) and a new tail.
This finishes Step~\(i\).
\end{itemize}
At the end of Step~\(i\) we have a network
\(\mathcal{N}_i\) whose vertices with in-degree~0
are in bijective correspondence with the clusters
in~\(\mathcal{C}_i\)
(see Figure~\ref{fig:process:non:binary:rtcn}(d) and~(e)).
\item
\underline{Step $n$:}
All vertices with in-degree~0 in the network obtained
after Step~$t-1$ are identified
as a single vertex which then forms the root~\(\rho\) of
the resulting network (see Figure~\ref{fig:process:non:binary:rtcn}(f)). 
\end{itemize}

Finally, each vertex in the rooted phylogenetic network
\(\mathcal{N} = (V,E,\rho)\) on~\(X\) generated
by the process described above is assigned a rank from
the set \(\{1,\dots,t\}\) (see Figure~\ref{fig:process:non:binary:rtcn}(f))
by putting:
\begin{itemize}
\item
\(\text{rank}(x)=1\) for all \(x \in X\),
\item
\(\text{rank}(u)=i\) for all vertices~\(u\) of the
network~\(\mathcal{N}_{i}\) obtained at the end of Step~\(i\)
such that~\(u\) is the head of an arc added in Step~\(i\)
(\(2 \leq i \leq t-1\)).
\item
\(\text{rank}(\rho) = t\).
\end{itemize}

We now summarize the key properties of the
rooted phylogenetic networks obtained by the process described
above which we will also call RTCNs.
The proof that these properties hold follows immediately from the
construction of the network~\(\mathcal{N}\) from the
given chain in~\((\mathfrak{T}(X),\preceq)\).

\begin{theorem}
\label{thm:non:binary}
For every chain \(\mathcal{C}_1,\dots,\mathcal{C}_t\)
in~\((\mathfrak{T}(X),\preceq)\) we obtain a rooted
phylogenetic network \(\mathcal{N}=(V,E,\rho)\) on~\(X\)
together with a map \({\rm rank} : V \rightarrow \{1,\dots,t\}\)
such that, for all \(1 \leq i < t\),
\[\mathcal{C}_i = \{C_v : {\rm there \ exists \ an \ arc} \ (u,v) \in E \
{\rm with} \ {\rm rank}(u) > i \geq {\rm rank}(v)\},\]
and \(\mathcal{C}_t = \{C_{\rho}\} = \{X\}\).
If \(t=n\) (i.e. the chain is a maximal chain on~\(X\)),
\(\mathcal{N}\)~is the binary RTCN that corresponds to the
chain by Theorem~\ref{thm:chain:encoding}.
\end{theorem}

Note that in the language of posets, this theorem implies that
the poset \((\mathfrak{T}(X),\preceq)\) is \emph{bounded}
because we have
\[\{\{x\}:x \in X\} \preceq \mathcal{C} \preceq \{X\}\]
for all \(\mathcal{C} \in \mathfrak{T}(X)\). This, together
with the fact that all maximal chains in \((\mathfrak{T}(X),\preceq)\)
have the same length, implies that \((\mathfrak{T}(X),\preceq)\)
is what is known as a \emph{graded} poset.
Moreover, Theorem~\ref{thm:gallery:connected} is
equivalent to saying that \((\mathfrak{T}(X),\preceq)\)
is \emph{gallery-connected}. Note that 
a similar relationship for nearest neighbor
interchanges on unrooted phylogenetic
trees on~\(X\) appears in~\citep{stadnyk2022edge}.

\begin{figure}
\centering
\includegraphics[scale=1.0]{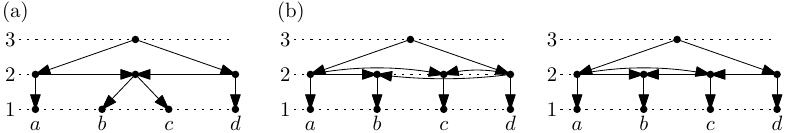}
\caption{(a) The non-binary RTCN on \(X=\{a,b,c,d\}\)
that we obtain by Theorem~\ref{thm:non:binary}
from the non-maximal chain
\(\mathcal{C}_1 = \{\{a\},\{b\},\{c\},\{d\}\}\)
\(\mathcal{C}_2 = \{\{a,b,c\},\{b,c,d\}\}\),
\(\mathcal{C}_3 = \{X\}\) in~\((\mathfrak{T}(X),\preceq)\).
(b) Two other non-binary tree-child networks with ranked
vertices that yield the same non-maximal chain.}
\label{fig:different:non:binary:rtcn}
\end{figure}

To conclude this section, 
we remark that Theorem~\ref{thm:non:binary}
only establishes that for each chain in the poset
\((\mathfrak{T}(X),\preceq)\) there \emph{exists} a suitable RTCN to 
represent this chain. Moreover, in case
the chain is maximal this RTCN is uniquely defined by
Theorem~\ref{thm:chain:encoding}.
However, for non-maximal chains there may be 
several different rooted phylogenetic networks
that are tree-child and have ranked vertices
(see Figure~\ref{fig:different:non:binary:rtcn}
for an example). This highlights  the fact mentioned
above that non-binary networks
are harder to capture. In particular, a more complex encoding would need to 
be devised if one wanted to define a metric  
on \emph{all} rooted phylogenetic networks
that are tree-child and have ranked vertices.
It could be interesting to explore this further in future work.

\section{Construction of a CAT(0)-orthant space of ETCNs}
\label{sec:consruction:network:space}

In this section, we define a distance on the 
collection of binary ETCNs having the same leaf set.
The main idea is to use the poset~\((\mathfrak{T}(X),\preceq)\)
introduced in Section~\ref{sec:non:binary:rtcn} to define
a continuous space of such networks and, by using properties of 
\((\mathfrak{T}(X),\preceq)\), 
show that this space is a so called CAT(0)-orthant space.

\begin{figure}[h]
\centering
\includegraphics[scale=1.0]{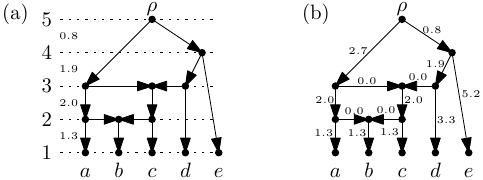}
\caption{(a) A binary RTCN on \(X=\{a,b,c,d,e\}\) where
positive real-valued differences between consecutive ranks are given.
(b) The corresponding equidistant weighting of the arcs of the
network.}
\label{fig:ex:etcn:cw}
\end{figure}

First, we need to present some more definitions. We call a
non-negative weighting of the
arcs in a binary RTCN~\(\mathcal{N}\) on~\(X\)
\emph{equidistant} if the total weight of the arcs in a directed path
in~\(\mathcal{N}\) from~\(\rho\) to some~\(x \in X\) 
does not depend on the choice of~\(x\) and the directed path
(see e.g. Figure~\ref{fig:ex:etcn:cw}). 
Note that, given non-negative real-valued differences
between consecutive ranks, an equidistant weighting is obtained 
by assigning to each arc the total difference between the rank of
its head and tail. Conversely, every equidistant weighting of
the arcs that is \emph{consistent} with the ranks of its vertices
(i.e. vertices of the same rank have the same distance from the root and the
higher the rank of a vertex the smaller the distance of it
from the root), clearly yields corresponding non-negative, real-valued
differences between consecutive ranks. Thus, to describe all
equidistant weightings of a binary RTCN that are consistent with
the ranks of its vertices, it suffices to look at all
possible ways to assign non-negative real-valued differences
between consecutive ranks.

\begin{figure}
\centering
\includegraphics[scale=1.0]{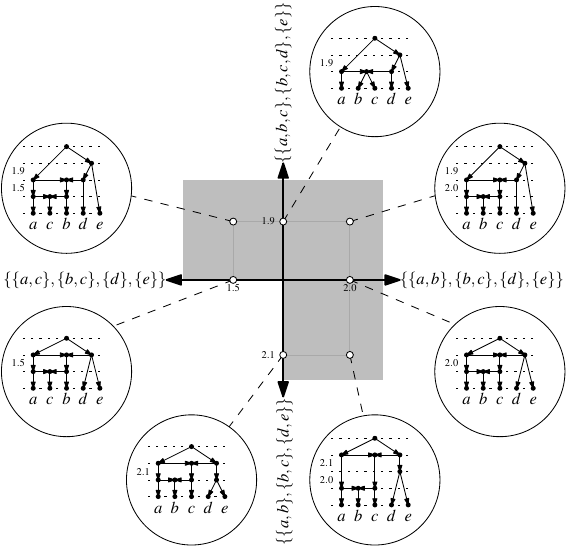}
\caption{The gray squares represent three orthants of maximum dimension
in the orthant space \(\mathfrak{S}(X)\) for \(X=\{a,b,c,d,e\}\).
These orthants are actually 4-dimensional. They are projected into 
the plane by showing only two of the four coordinate axes that
determine each of them (each coordinate axis is labeled
by a cluster system in~\(\mathfrak{T}(X)\); the two axes
corresponding to the cluster systems \(\{\{a\},\{b\},\{c\},\{d\},\{e\}\}\)
and \(\{\{a,b,c\},\{b,c,d,e\}\}\) are not shown in the projection).
Each point in an orthant corresponds to
an ETCN on~\(X\) with the coordinates of the point
describing the difference between consecutive ranks.}
\label{fig:ex:orthants}
\end{figure}

To make this more precise, we use again the fact that,
by Theorem~\ref{thm:chain:encoding},
binary RTCNs on~\(X\) are in bijective correspondence
with maximal chains 
\[\{\{x\}:x \in X\}=\mathcal{C}_1,\dots,\mathcal{C}_n=\{X\}\]
on~\(X\). Assigning positive, real-valued differences
between consecutive ranks then corresponds to a map~\(\omega\)
that assigns, for all \(1 \leq i < n\),  to the
cluster system~\(\mathcal{C}_i\) a
positive real number~\(\omega(\mathcal{C}_i)\). To illustrate this,
consider again the example in Figure~\ref{fig:ex:etcn:cw}(a), where
we obtain the following map \(\omega\):
\begin{align*}
\omega(\mathcal{C}_1) &= \omega(\{\{a\},\{b\},\{c\},\{d\},\{e\}\}) = 1.3,\\
\omega(\mathcal{C}_2) &= \omega(\{\{a,b\},\{b,c\},\{d\},\{e\}\}) = 2.0,\\
\omega(\mathcal{C}_3) &= \omega(\{\{a,b,c\},\{b,c,d\},\{e\}\}) = 1.9,\\
\omega(\mathcal{C}_4) &= \omega(\{\{a,b,c\},\{b,c,d,e\}\}) = 0.8.
\end{align*}
The maps~\(\omega\) for a fixed binary RTCN form an
\((n-1)\)-dimensional \emph{orthant} in $\mathbb{R}^{(n-1)}$ that is spanned by
the \(n-1\) axes that each correspond to one of the
cluster systems \(\mathcal{C}_1,\dots,\mathcal{C}_{n-1}\).
For example, the orthant for the binary RTCN in Figure~\ref{fig:ex:etcn:cw}(a)
is illustrated in Figure~\ref{fig:ex:orthants} along with the
orthants for two other binary RTCNs.
Orthants for different binary RTCNs may share some of their 
axes. This is the case precisely when the corresponding
maximal chains on~\(X\) share some of its cluster systems.
Intuitively, as can be seen in Figure~\ref{fig:ex:orthants},
orthants are ``glued'' together along these shared axes
and, in this way, we obtain a continuous space whose points
are meant to represent binary RTCNs on~\(X\) with an equidistant
weighting of its arcs that is consistent with
the ranks of the vertices.

One technical aspect, however,
also illustrated in Figure~\ref{fig:ex:orthants}, is that
points that lie on the boundary of an orthant correspond to
maps~\(\omega\) that assign~\(0\) to certain cluster
systems. Intuitively, this means that the cluster system
is skipped, leading to a (non-maximal)
chain in the poset~\((\mathfrak{T}(X),\preceq)\) 
which then corresponds to a (not necessarily binary) RTCN on~\(X\)
obtained by Theorem~\ref{thm:non:binary}.
In view of this, we call a (not necessarily binary) RTCN~\(\mathcal{N}\)
obtained by Theorem~\ref{thm:non:binary} together with
an equidistant weighting of the arcs of~\(\mathcal{N}\)
that is consistent with the ranks of the vertices of~\(\mathcal{N}\)
an \emph{equidistant tree-child network} (ETCN) on~\(X\).

A concise, formal description of the continuous space
we have just described can be obtained by considering
maps \(\omega : \mathfrak{T}(X)-\{\{X\}\}  \rightarrow \mathbb{R}_{\geq 0}\).
For such a map, put \(\supp(\omega)
= \{\mathcal{C} \in \mathfrak{T}(X) : \omega(\mathcal{C}) > 0\}.\)
Then the \emph{orthant-space} \(\mathfrak{S}(X)\)
of all ETCNs on~\(X\) consists of all maps
\(\omega : \mathfrak{T}(X)-\{\{X\}\} \rightarrow \mathbb{R}_{\geq 0}\)
such that the cluster systems in \(\supp(\omega)\)
(together with the cluster systems \(\{\{x\}:x \in X\}\) and \(\{X\}\)),
when ordered by~\(\preceq\),
form a chain in the poset~\((\mathfrak{T}(X),\preceq)\).
More details about the general construction of an
orthant-space based on the chains in a poset can be
found in~\citep[Sec.~4.1]{huber2024space}.
We remark that this construction can also be
used to obtain the space of ultrametric trees
presented by~\citet{GD16} (cf. \citet{huber2024space}).

We now show that the space
\(\mathfrak{S}(X)\) comes equipped with a distance
that has some attractive properties.
More specifically, in the theorem below we show that \(\mathfrak{S}(X)\)
together with the distance~\(\delta\) that assigns the
length \(\delta(\omega,\omega')\) of a shortest 
path\footnote{A path is essentially a connected, finite sequence of straight line segments,
	and the length of a path is the sum of the Euclidean lengths
	of each of the line segments.} or \emph{geodesic} between
any two points \(\omega, \omega'\) in \(\mathfrak{S}(X)\) 
is a \emph{CAT(0)-orthant space}. Note that this 
immediately implies that
there is a unique geodesic between any two points in \(\mathfrak{S}(X)\).   
As it is quite technical and not important for the proof, we shall not present the definition of 
CAT(0)-orthant spaces here, but instead refer the reader to e.g. \cite[Section 6]{MOP15a} for more details.

\begin{theorem}
\label{thm:etcn:space}
The metric space \((\mathfrak{S}(X),\delta)\)
is a CAT(0)-orthant space whose points are in bijective
correspondence with ETCNs on~\(X\).
\end{theorem}

\pf
It is known (see e.g. \citet[Sec.~4.1]{huber2024space} for more
details), that constructing a metric space based on a poset in the
way that \((\mathfrak{S}(X),\delta)\) was constructed based on the
poset~\((\mathfrak{T}(X),\preceq)\) always yields a CAT(0)-orthant space.

We now show that the points in \((\mathfrak{S}(X),\delta)\) are in bijective
correspondence with ETCNs on~\(X\).
First note that each point \(\omega \in \mathfrak{S}(X)\) corresponds to 
a chain \(\mathcal{C}_1, \dots , \mathcal{C}_t\)
in \((\mathfrak{T}(X),\preceq)\) that is obtained
by ordering the cluster systems in~\(\supp(\omega)\)
together with the cluster systems \(\{\{x\}:x \in X\}\) and \(\{X\}\)
by~\(\preceq\). By Theorem~\ref{thm:non:binary}, the
chain yields a well-defined (but not necessarily binary) RTCN on~\(X\).
From the values~\(\omega(\mathcal{C}_i)\), \(1 \leq i < t\),
we obtain an equidistant weighting of the arcs of this RTCN
that is consistent with the ranks of the vertices
as described in this section.

Conversely, assume we are given an ETCN on~\(X\), that is,
a (not necessarily binary) RTCN~\(\mathcal{N}\) on~\(X\)
together with an equidistant weighting of the
arcs that is consistent with the ranks of the vertices of~\(\mathcal{N}\).
Let \(\mathcal{C}_1, \dots , \mathcal{C}_t\) be the chain
in \((\mathfrak{T}(X),\preceq)\) that corresponds to~\(\mathcal{N}\)
by Theorem~\ref{thm:non:binary}. As described in the text,
the given equidistant weighting of the arcs of~\(\mathcal{N}\)
yields non-negative values \(\omega(\mathcal{C}_i)\) for
all \(1 \leq i < t\). We formally extend these to a map
\(\omega: \mathfrak{T}(X) - \{\{X\}\} \rightarrow \mathbb{R}_{\geq 0}\) by
putting \(\omega(\mathcal{C})=0\) for all
\(\mathcal{C} \in \mathfrak{T}(X) 
- \{\mathcal{C}_1, \dots , \mathcal{C}_{t-1},\{X\}\}\),
which then yields the point in \(\mathfrak{S}(X)\) corresponding
to the given~ETCN.
\epf

As an immediate corollary of Theorem~\ref{thm:etcn:space},
it follows that the distance between any two
ETCNs on~\(X\), that is, the value \(\delta(\omega,\omega')\)
for the corresponding maps \(\omega,\omega' \in \mathfrak{S}(X)\),
can be computed in polynomial time~\citep[Corollary~6.19]{MOP15a}.

\section{Conclusion}
\label{sec:conclusion}

In this paper, we have presented various ways to compare binary RTCNs.
Interestingly, it is shown by~\citet{collienne2021computing}
that, given two binary ranked trees \(\mathcal{T}_1\) and
\(\mathcal{T}_2\) on~\(X\), the rNNI-distance
between \(\mathcal{T}_1\) and \(\mathcal{T}_2\) 
can be computed in polynomial time.
It would be nice to know if the analogous
rNNI$^*$-distance between two binary RTCNs
defined in Section~\ref{sec:nni:on:rtcn} can also
be computed in polynomial time.
In this vein, it might also be of
interest to consider alternative distances on RTCNs that might arise from
generalizing other types of ranked tree modifications
(for example, {\em subtree prune and regraft operations
(SPRs)} considered by~\citet{collienne2024ranked}),
or to investigate how the ranked tree distances
considered by~\citet{kim2020distance} might be generalized to RTCNs.

In another direction, it could be worth investigating 
combinatorial properties of the poset \((\mathfrak{T}(X),\preceq)\).
For example, we have shown that this poset is 
gallery-connected, a property that, for \emph{any} finite poset,
is immediately implied in case the poset is {\em shellable} 
(see e.g. \citet{BW83} for a formal definition of shellability).
Is \((\mathfrak{T}(X),\preceq)\) shellable? 
If this were true, then it would immediately 
imply that the space \((\mathfrak{S}(X),\delta)\)
considered in Section~\ref{sec:consruction:network:space} 
has some special topological properties.
Note that a similar combinatorial technique was used by~\citet{AK06a}
to understand the topology of spaces of (unranked) equidistant trees.

Finally, Theorem~\ref{thm:etcn:space} implies that 
methods for performing a variety of statistical
computations (e.g. Fr\'echet mean and
variance~\citep{bacak2014computing,MOP15a},
an analogue of partial principal component analysis~\citep{nye2017principal}
and confidence sets \citep{willis2019confidence})
can be applied (or extended) to the 
metric space \((\mathfrak{S}(X),\delta)\).  
These methods allow, for example, the computation of a \emph{consensus} for a 
collection of ETCNs. It would be interesting to further explore this
possibility, and also to investigate geometric properties of the space \((\mathfrak{S}(X),\delta)\).\\



\bibliographystyle{apalike}
\bibliography{network}

\end{document}